# Multi-element cylindrical electrostatic lens systems for focusing and controlling charged particles

**Omer Sise, Melike Ulu and Mevlut Dogan**

Afyon Kocatepe University, Science and Art Faculty, Department of Physics, Afyon, 03200 Turkey

**Abstract**

This paper describes theoretical modelling of electrostatic lenses based on 3, 4 and 5 closely spaced cylindrical electrodes, respectively. In each case, modelling is carried out numerically using commercial packages SIMION and LENSYS, and a variety of performance parameters are obtained. These include the magnification, the $3^{rd}$ order spherical and chromatic aberration coefficients. Special cases such as zoom lens (i.e., lenses whose magnification may be changed without losing focus) are considered. Results are obtained as a function of the ratios of the electrode lengths and gaps, and as a function of ratios of the controlling voltages.

As a result, it is shown that how a multi-element lens system can be operated with the whole focal properties in a useful mode for using in experimental studies.

## 1. Introduction

Electrostatic lenses are widely used to control beams of charged particle with various energy and directions in several fields, especially in electron spectroscopy. Therefore their focal properties have been extensively studied theoretically and also experimentally. It is well known that a two-element lens system cannot keep the image position constant while varying the ratio of final to initial electron energy. For that reason, it is necessary to add at least one further element to the system. The three-element lens systems can be operated while keeping the image position constant. Two-element electrostatic lenses have been used in low-energy electron spectrometers to increase sensitivity and resolution [1-3]. Details of the properties several types of two-element lenses can be found in Refs. [4-14]. The focal properties of three-element lenses with zoom type of optics have been obtained by Refs. [15-22] for various diameters and voltage ratios. One of the types of the three-element lens is as an "einzel" lens in which the outer electrode are held at the same potential and beam focusing is achieved by varying the potential of the centre electrode. Einzel (unipotential) lenses are necessary when a beam is to be focused without changing its energy [23-27]. The condition of higher energy at the centre of three-element lenses results in a lower aberration coefficient of the image and this mode of operation is preferred. Details of aberration coefficients have been determined in Refs. [28-30].

Several zoom lenses which consist of four elements are used to ensure that focusing in the monochromator and analyser optics would be independent of the electron kinetic energy [31, 32]. Therefore significant effort in the past has been made to determine the focal properties of these lenses. Although multi-element lenses are expected to be more flexible and allow controlling of many independent lens parameters, two and three-element lenses have been studied more often than the four-element lenses. The four-element lens systems have been investigated experimentally [33] and theoretically [34, 35], and showed that four element lenses are necessary to produce an image at a specific position and magnification. An extensive study including spherical aberration effects has been described in Refs. [36] and [37] for a variety configurations of four-element lens systems having A/D=0.5 where A is the length of the centre electrodes including half the gap to each side. A four-element lens with zoom type which described by Ref. [37] has been used recently in the low energy electron-molecule scattering experiments to provide constant image position and magnification [38].

Böker *et al* [39] have presented several different configurations of four-element lens systems operated in angular resolving mode. Apart from this dimension, it is also possible to operate these lens systems for A/D=1. To our knowledge, this mode has only been mentioned in Ref. [40] but no detailed study has been performed yet. We present the focal properties of the four-element lenses having A/D=0.5 and 1 with G/D=0.1, and compare the characteristic properties of these two different lens systems.

Based on the concerns described above, in an electrostatic lens system quantitative information is required over a wide energy range and a zoom-type of optics is needed. If the magnification is to remain constant over a wide range of energies, quite complicated electrostatic lens systems are required, containing three, four, or even more lens elements. Three element lenses can be used to aid the design of lenses having fixed image position, but the magnification is not constant. However, the four element lens systems can be operated with a constant magnification while keeping the image position also constant. Electrostatic lens systems with more than four elements are generally used in experimental studies [43, 44] to maintain a truly zoom lens with constant magnification overall voltage ratios. Additionally five and more element electrostatic lenses can be operated in the afocal mode which rays incident parallel to the axis leave the lens system still parallel to the axis [45, 46]. The focal properties of such multi-element lenses are derived from calculated data of three and four element lenses, especially for afocal lenses. In the literature, there have been mainly two previous studies on the five element lens systems. An experimental study of a five element lens was made by Heddle and Papadovassilakis [47] for afocal lens which described by Heddle [45] and for true-zoom lens, and Trager-Cowan *et al* [48] described the behaviour of two five-element lens systems by using standard matrix methods and calculated focal properties and spherical aberration coefficients of these lenses. They have concentrated on a five-element lens system having A/D=1.5, and calculated focusing properties of the lens system. We present the results of a simple analysis of the five-element lens system having A/D=0.5 and 1 with the help of SIMION and LENSYS (Varimag 6 and 8 procedures) programs. Therefore we review the focal and magnification properties of multi-element cylindrical electrostatic lenses in great detail. Special cases of lenses, such as zoom lens, were also discussed. In addition, voltage ratios, spherical and chromatic aberration coefficients of the resulting zoom lens have been evaluated. So, there are much more data in the following, but it is necessary for this investigation.

The novelty in this paper is limited, since lenses of these types have existed since the early days of electron microscopy. However, there may be scope for accurate numerical modelling of these structures using computer-based techniques that were not available to earlier workers.

## 2. Determination of axial potential

Calculation of the properties of electrostatic lens systems required knowledge of the potential function, V(r), along the whole optical axis. There are four general methods to solve the potential distribution. One of these is the Separation of Variables Method (SVM) which is used to solve Laplace's equation with cylindrical symmetry. In this method the solution of Laplace's equation is written as a number of functions where each function depends on one variable only [9, 15, 16, 18, 24, 49-51]. With the Boundary Element Method (BEM), or Charge Density Method (CDM), the system of electrostatic lenses is replaced under applied potentials with a system of rings of charge that assumes the same geometry as the cylinders [11, 14, 17, 52-55]. The Finite Element Method (FEM) is a numerical technique for obtaining solutions to boundary value problems. This method is used widely to obtain the field distribution. The principle of the method is that the potential distribution will be such that the potential energy of the electrostatic field is a minimum and so this potential energy is expresses in terms of the potentials at the vertices of each and all the triangular elements [56-58]. A solution of Laplace equation can be found by the Finite Difference Method (FDM). Among the most often used are the five-point and nine-point relaxation techniques [59-62].

In this paper we have been assisted in determining focal properties of the electrostatic lens systems by SIMION and LENSYS which is the FDM programs to obtain potential distribution and allow visualising the trajectories of charged particles. A comparison of the accuracy and speed of FDM programs and BEM programs (i.e. CPO [63]) based on these methods can be found in Ref. [64].

## 3. Parameterisation of electrostatic lenses

The focal properties of a lens can be described by several parameters. Figure 1 shows the geometry and the focal points of three-element lens which consist of three coaxial cylinders of the same diameter D, separated by a distance G. The cylinders are at potentials $V_1$, $V_2$ and $V_3$. $F_1$ and $F_2$ are mid-focal lengths and the parameters $f_1$ and $f_2$ are the first and second focal lengths. The relationships between lens parameters are related through the equation (1) where i is the number of lens elements [42]. Other parameters are aberration coefficients and one type of these is the spherical aberration coefficient ($C_S$), where the outer zones of the lens focus more strongly than the inner zones. Another type is chromatic aberration coefficient ($C_C$), where charged particles of slightly different energies get focused at different image plane. Both types of aberrations can be minimised by reducing the convergence angle of the system so that the charged particles are confined to the centre of the lenses.

$$M = -\frac{f_1}{P - F_1} = -\frac{Q - F_2}{f_2} \qquad f_2 = \sqrt{\frac{V_i}{V_1}}\, f_1 \qquad (1)$$

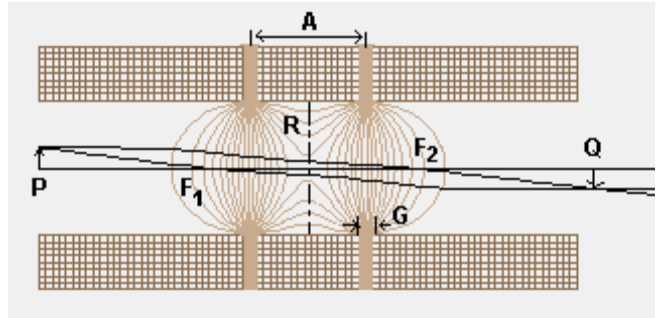

**Figure 1.** A three-element lens system with the equipotential surfaces and obtaining the focal properties for a given voltage ratios.

## 4. Results and discussions

4.1. Three-element lenses

*4.1.1. The length of the centre electrode*

Focal properties of a lens depend on both lens geometry and the voltage ratios. Especially, the length of the centre element A/D affects the behaviour of both the magnification and aberration coefficients. For this reason, it is necessary to determine the most appropriate dimensions of the lens system.

Three-element lenses are widely used to satisfy the condition of maintaining a zoom lens. One needs to know the relationship between $V_2/V_1$ and $V_3/V_1$, for fixed values of the object and the image distance P/D and Q/D. The calculated values of some zoom lenses are shown in figure 2, for three values of A/D=0.5, 1 and 2, for the fixed value of P/D and Q/D, 3.5 and 2.5 respectively. It is shown that as $V_3/V_1$ is varied, the magnification may also vary, and the range of the magnification for A/D=0.5 is less than for A/D=1 and 2. It is shown that this range is 0.4-1 for A/D=0.5 and 0.2-2.2 for A/D=2.

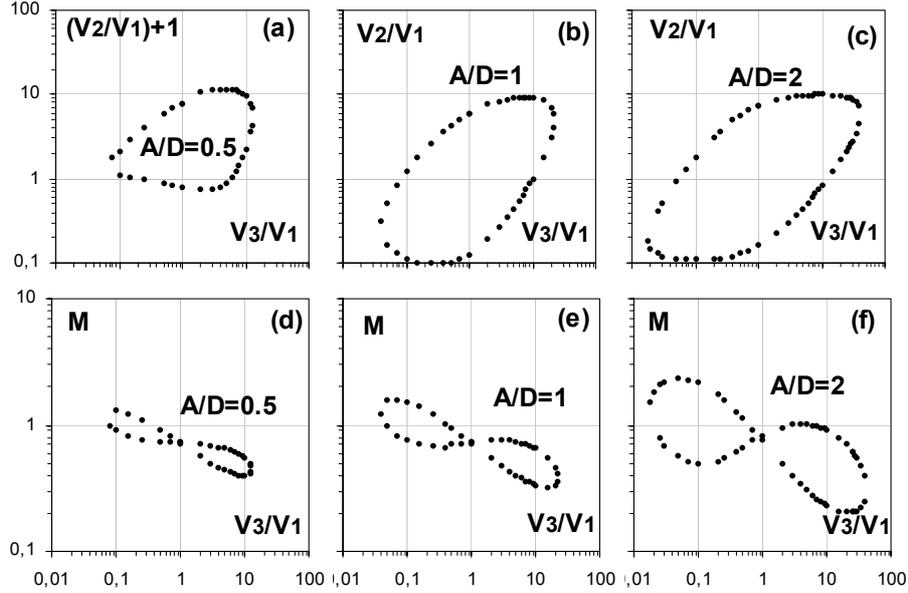

**Figure 2.** Zoom lens curves for various the length of the centre electrode in three-element lenses having constant object and image distance, P/D=3.5 and Q/D=2.5 for A/D=0.5, 1 and 2, respectively. Note that the scale of $V_2/V_1$ in (a) is shifted one unit.

*4.1.2. Relationship between voltage ratios*

In this part we consider only a three element lens having A/D=1. Lens parameters have been calculated over large ranges of values $V_3/V_1$ and $V_2/V_1$ for accelerating ($V_3/V_1>1$) and decelerating ($V_3/V_1<1$) mode and are given in figure 3. The second focal length, $f_2/D$ can be found equation (1). It can be seen that as $V_2/V_1$ increases the focal lengths tend to become shorter and to be less dependent on $V_3/V_1$. In addition to the lens parameters, the spherical and chromatic aberration coefficients have been calculated (but not shown) for both accelerating and decelerating lenses. The aberration coefficients, $C_s/D$ and $C_c/D$, for three-element lenses with the higher values $V_2/V_1$ are always less than the lower values of $V_2/V_1$. Because the paths of charged particles with the higher values of $V_2/V_1$ in the centre of the lens is much closer to the axis. $C_s$ and $C_c$ differ very fast for a given P at the small values of $V_2/V_1$ and the accelerating lens has a lower coefficient of spherical and chromatic aberration than the decelerating lens.

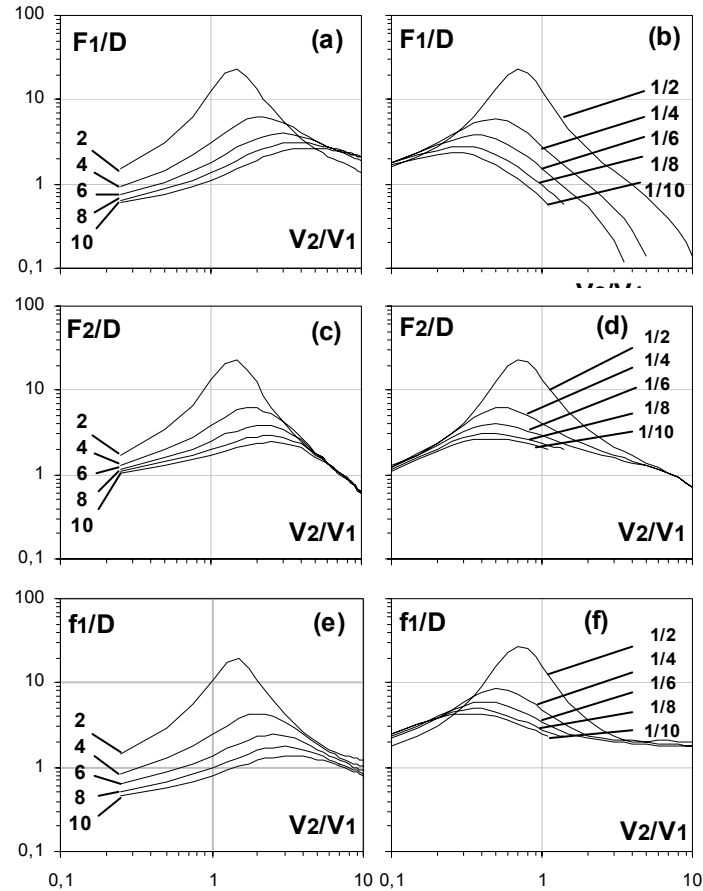

**Figure 3.** Calculated values of the lens parameters for acceleration ($V_3/V_1$=2, 4, 6, 8 and 10) and deceleration ($V_3/V_1$=1/2, 1/4, 1/6, 1/8 and 1/10) lenses having A/D=1 as a functions of $V_2/V_1$ for various ratios of $V_3/V_1$ (on the curves).

*4.1.3. Three- element zoom lens*

It is possible to use three-element lenses as zoom lenses. We should know the relationship between $V_2/V_1$ and $V_3/V_1$ for fixed values of P, Q and A. Figure 4 shows these relationship for three different values of Q/D (2.8, 4 and 10) and the variation of the magnification of the image distance for P/D=3.5 and A/D=1. On each zoom lens curve for $V_3/V_1$=1 two points correspond to an einzel lens and for $V_2/V_1$=1 or $V_2/V_1$=$V_3/V_1$ four points correspond to a two-element lens.

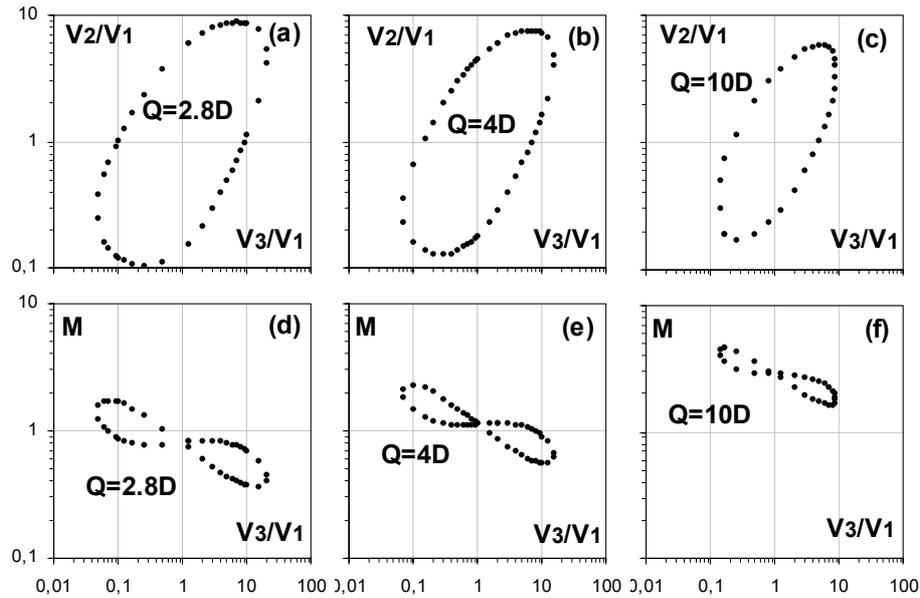

**Figure 4.** Zoom lens curves for various fixed image distance for G/D=0.1 and A/D=1. Showing the relationship between $V_2/V_1$ and $V_3/V_1$ and the variation of the magnification M with $V_3/V_1$ for Q/D=2.8, 4 and 10, respectively.

The zoom lens condition for three-element lens having the potentials of the second and third electrodes fixed may be achieved with each of the potentials at a high and low value $V_2/V_1$ and $V_3/V_1$. Figure 5 illustrates the zoom lens properties of the lens in a different way. If the lens is operated with $V_2/V_3>1$ for high mode the magnification M changes by a small amount from 0.74 to 0.88 for Q/D=2.8 when the ratio of final to initial energy ($V_3/V_1$) is changed from 8 to 0.1, where $V_2$ is altered to keep the image position constant. The four possible configurations lead to significantly different coefficients of spherical and chromatic aberration for all values of the magnification. Calculated values of the spherical and the chromatic aberration coefficient for a range of values of $V_2/V_1$ and $V_3/V_1$ are also shown. With its low spherical and chromatic aberration and small change in magnification this is obviously the better mode in which to operate the zoom lens as a variable acceleration system. If it is desired to vary magnification then the outer mode of operation in which the centre electrode is at a 'Low' voltage can be used, although this mode has inherently higher spherical and chromatic aberration. It was shown that for all values of the magnification, the spherical and the chromatic aberration coefficients were least for $V_2/V_1>1$ and $V_3/V_1>1$, that is 'High' mode.

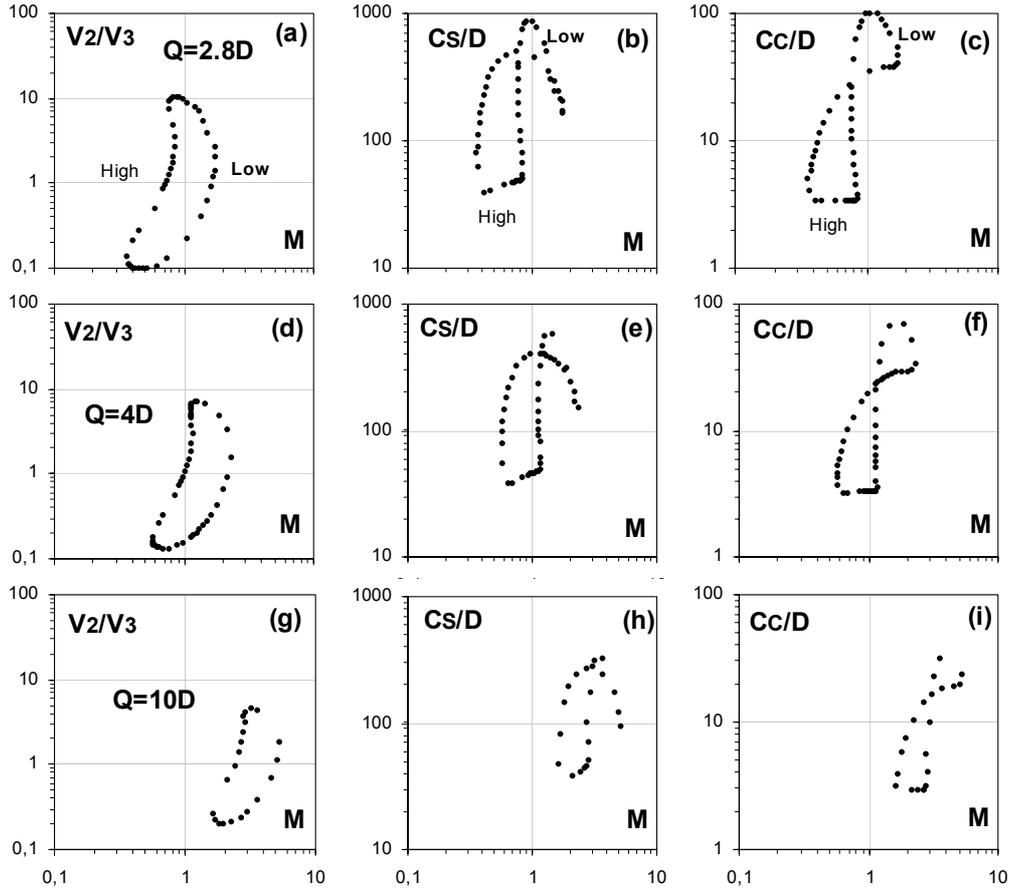

**Figure 5.** The variation of $V_2/V_3$ which contains two voltage ratios, the spherical and chromatic aberration coefficient, $C_s/D$ and $C_c/D$, of the three-element zoom lens as a function of the magnification for values of $V_2/V_1$ and $V_3/V_1$ at "High" and "Low" mode with the various image distance $Q/D$, 2.8 (a)-(c), 4 (d)-(f) and 10 (g)-(i).

4.2. Four element lenses

It is possible to use four-element lenses as zoom lenses and the energy of the image and its magnification are constant during energy changing. It could be an advantage to have a lens which could produce an image at a fixed position and magnification. In fact this is probably most important for use in experimental studies.

4.2.1. Focal properties

Martinez and Sancho [36] have presented the optical properties of the four-element lens having $A/D=0.5$ by using the charge density method. In this work we consider the lens system with $A/D=1$ and our simulation programs use the finite difference method to calculate the potential distribution. With these programs we made the similar investigation of treating the lenses and compared the focal and zoom lens properties of these two lens systems which have different dimensions.

Calculated values of the first and second mid-focal lengths $F_1/D$ and $F_2/D$ are shown in figures 6 and 7, for a range of values $V_2/V_1$ from 0.1 to 10, for five different values of $V_3/V_1$ from 0.5 to 5, and for four different values of $V_4/V_1$ from 1 to 10.

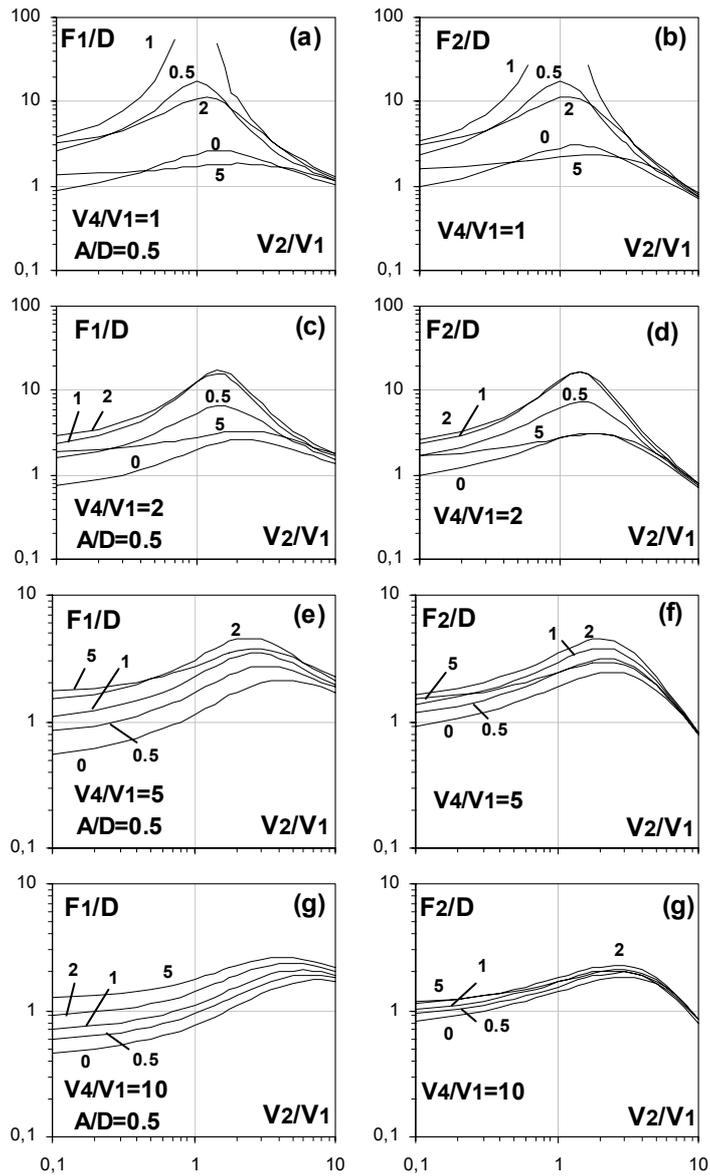

**Figure 6.** Values of the first and second mid-focal length $F_1/D$ and $F_2/D$ for four-element lenses with $A/D=0.5$ which was investigated by Ref. [36] as a function of $V_2/V_1$ for different values of $V_3/V_1$ (on the curves) and $V_4/V_1$.

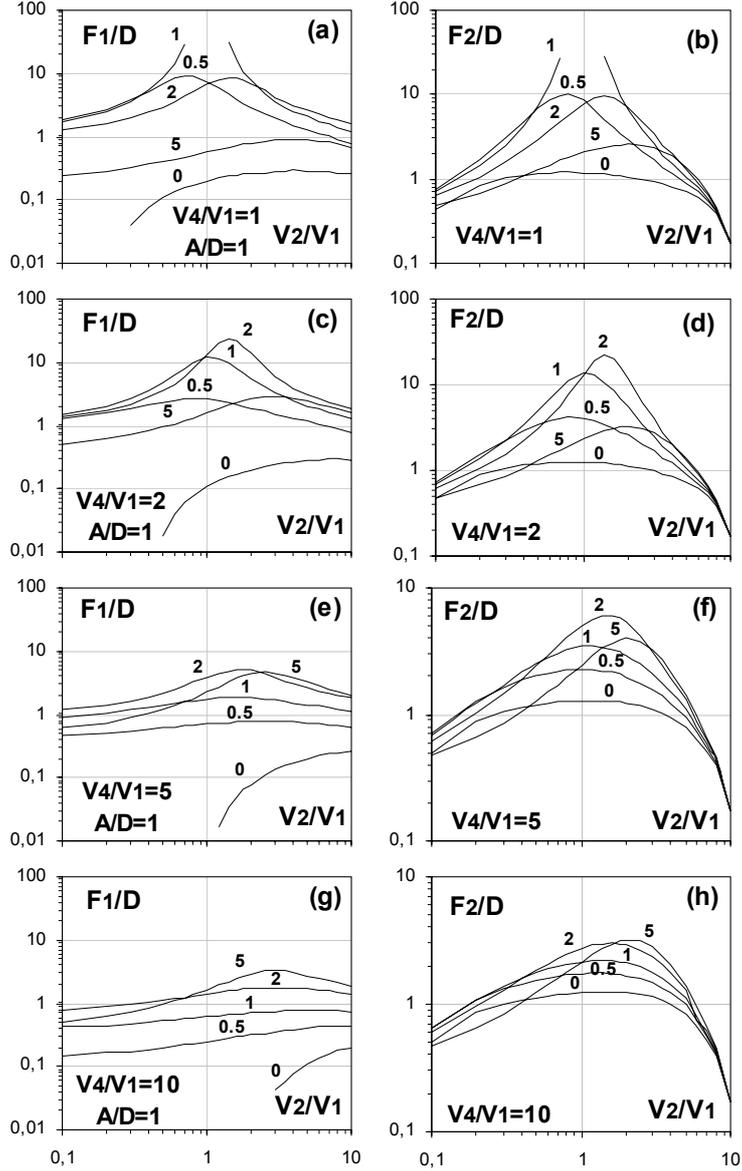

**Figure 7.** Values of the first and second mid-focal length $F_1/D$ and $F_2/D$ for four-element lenses with $A/D=1$ as a function of $V_2/V_1$ for different values of $V_3/V_1$ (on the curves) and $V_4/V_1$.

The mid-focal lengths, $F_1/D$ and $F_2/D$, have the same values when the value of $V_3/V_1$ and $V_4/V_1$ is near to 1.0, but as $V_3/V_1$ diverges from unity the four-element lens becomes stronger. On the other hand since the potential of the final element $V_4/V_1$ is taken high values, the lens parameters are independent of the voltage fluctuations of $V_3/V_1$. These results are very similar to the results of the previous work [36].

4.2.2. Fixed image position and magnification

For a zoom lens the dependence of M and the aberration coefficients on the ratio $V_3/V_2$ for a given overall voltage can be useful design guide and it would be instructive to present the data in this way. The relationship between $V_2/V_1$ and $V_3/V_1$, for fixed values of $P/D=2$, and $V_4/V_1=5$, for either Q is constant, or M is constant, is shown in figure 8, for two different values of A/D. In order to test the present method, we have calculated the focal properties of the four-element lens with $A/D=0.5$ and $G/D=0.1$. A few of the obtained results are compared to the previous ones (broken curve) in figure 8(a), when the image position is constant. It could be seen that a very good agreement with the previous data was achieved.

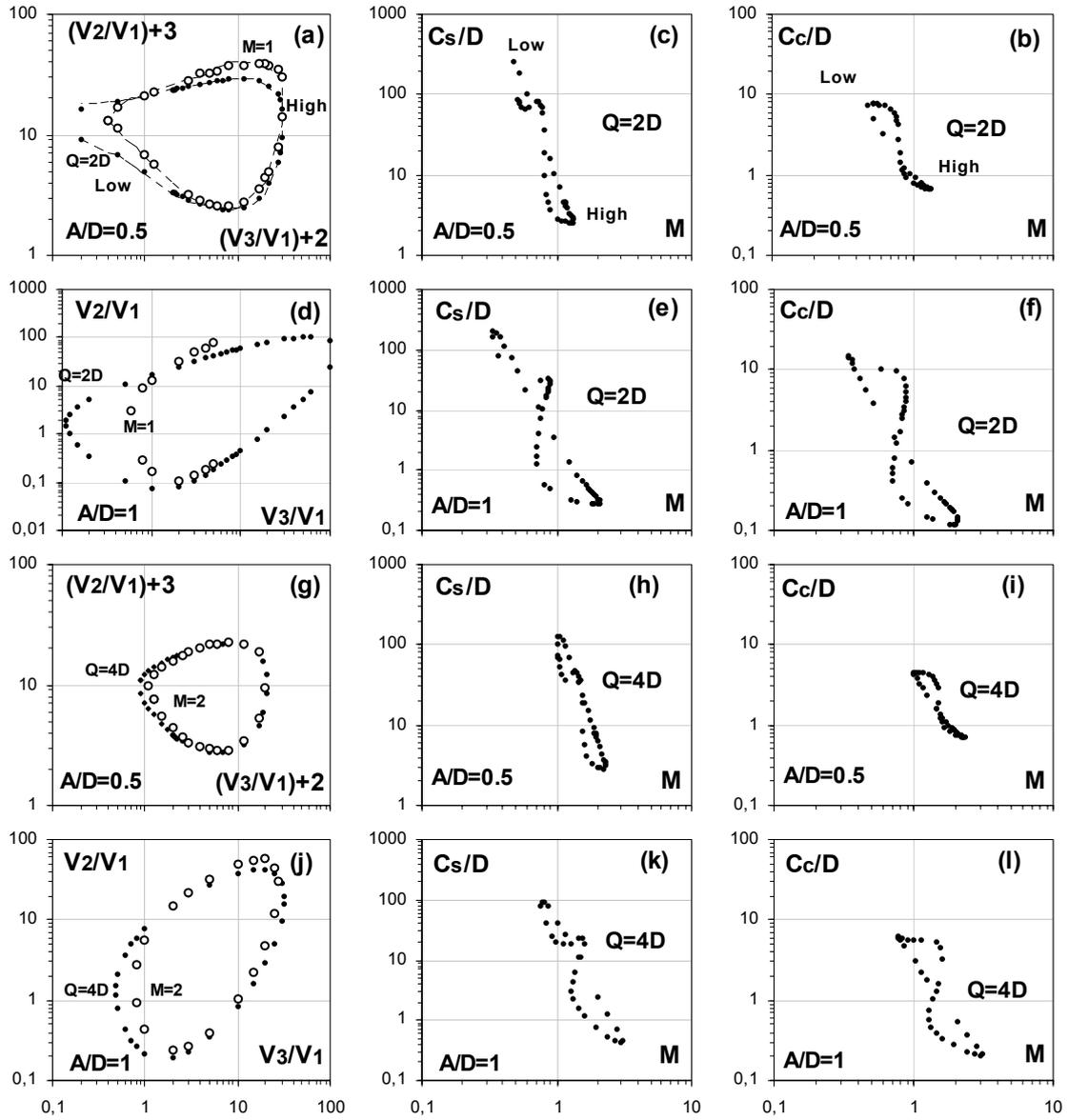

**Figure 8.** Zoom lens curves of four-element lenses having A/D=0.5 and 1 for P/D=2 and $V_4/V_1$=5, for either Q is constant (●) or M is constant (○) (broken curve in (a); Ref. [37]).

If the lens is operated with different cases it can be seen that for Q/D=2 the magnification M changes about 1.0 and for Q/D=4 changes about 2.0. Then we are able to determine the most appropriate values of M for a given P and Q.

It can be seen that there are two crossing points which corresponding to Q=2, M=1 and two corresponding to Q/D=4, M=2 for two different values of A/D. For other values of $V_4/V_1$ we will determine the relationship between voltage ratios for several combinations of P, Q and M values. However for A/D=2 the magnification of the lens begins to increase very strongly after particular values of voltage ratio $V_2/V_1$. Because the image position coincides with the lens gap between the third and the final element of the lens where act of deviating of the lens is very strong and the magnification takes highest values. Then the values of $V_2/V_1$ begin to increase to provide constant magnification at this region for M=1.

Calculated values of the spherical and the chromatic aberration coefficient of resulting zoom lens for a range of values of $V_2/V_1$ and $V_3/V_1$ are also shown. 'High' mode, which the values of $V_2/V_1$ and $V_3/V_1$ are at a high voltage, has the lowest spherical and chromatic aberration coefficients and this is the better mode for using four-element lenses as a zoom lens. Other cases have inherently higher aberration coefficients, essentially at the lower voltages ('Low' mode). Of the two four-element lenses we have studied, the lens having A/D=0.5 offers the advantage of a slightly wider range of magnifications compared with A/D=1.

When we compare the spherical and chromatic aberrations for two different values of A/D, it could be seen that the aberration coefficients for A/D=1 is lower than A/D=0.5 for the same values of the lens parameters. For example the minimum values of $C_s$ for Q=2 are 1.5 and 0.2, and the minimum values of $C_c$ are 0.62 and 0.11 for A/D=0.5 and 1, respectively. But for Q=2 the range of the magnification for A/D=0.5 is from 0.46 to 1.2, and for A/D=1 is from 0.32 to 2.05.

Similar zoom curves are applied for other values of P and Q to see the effect of the object and image distance on the range of the working voltage ratios. But the magnification must be taken different from previous values because the necessary values of the voltage ratios for Q-M pairs do not match for different object distances P.

### 4.2.3. *Constant value of P, Q and M*

We calculated the focal parameters of four-element lens systems overall voltage ratios. Now we chose six combinations for A/D=0.5 and 1 so that the object position P, image position Q and magnification M are constant. This configuration will be denoted as P2Q2M1. The properties of six four-element lens configurations for each type of lens system are presented here.

There are several crossing points corresponding to constant P, Q and M values for $V_4/V_1$=5 in figure 8. For other values of $V_4/V_1$ we can determine the necessary values of $V_2/V_1$ and $V_3/V_1$ for combinations of P, Q and M values. Martinez *et al* [37] have chosen the most appropriate combinations of values of P, Q, and M, and presented results for A/D=0.5. In this paper we have used their six different combinations, and applied this parameterisation to the four-element lens having A/D=1 with $P^{(1)} = P^{(0.5)} + 1$ and $Q^{(1)} = Q^{(0.5)} + 1$ where the superscript illustrates the value of A/D. Ref. [37] have presented the relationship between voltage ratios $V_2/V_1$, $V_3/V_1$ and $V_4/V_1$ for only two combinations. We calculated results for other combinations for two types of four-element lens system and compared the interval of the voltage ratio $V_4/V_1$. In addition, the first mid-focal lengths, the spherical and chromatic aberration coefficients have been evaluated as a function of $V_4/V_1$ for each number of P, Q and M values. This design of figures 9 and 11 might help for charged particle experiments [38].

The values of the fixed lens parameters for six combinations are given in table 1 and the relationship between $V_2/V_1$ and $V_3/V_1$ as a function of $V_4/V_1$ for fixed values of P, Q and M are shown in figure 9 for two different lens types. The range of the values of $V_4/V_1$ includes the magnification is equal to 1.0 for four combinations and the ratio of $V_4/V_1$ is greater than 15 for three combinations. These latter conditions are especially convenient for electron impact studies.

**Table 1.** Fixed lens parameters for the six combinations.

|  | No | 1 | 2 | 3 | 4 | 5 | 6 |
|---|---|---|---|---|---|---|---|
| Ref. [37] | A/D=0.5 | P2Q2M1 | P2Q3M1 | P2Q4M1 | P2Q4M2 | P3Q3M1 | P4Q2M0.5 |
|  | $V_4/V_1$ | 1.0-29.0 | 2.0-30.0 | 6.0-32.0 | 1.0-14.5 | 1.0-7.0 | 1.0-6.8 |
| This work | A/D=1 | P3Q3M1 | P3Q4M1 | P3Q5M1 | P3Q5M2 | P4Q4M1 | P5Q3M0.5 |
|  | $V_4/V_1$ | 1.0-23.0 | 1.0-30.0 | 7.0-30.0 | 1.0-9.0 | 1.0-10.0 | 1.0-15.0 |

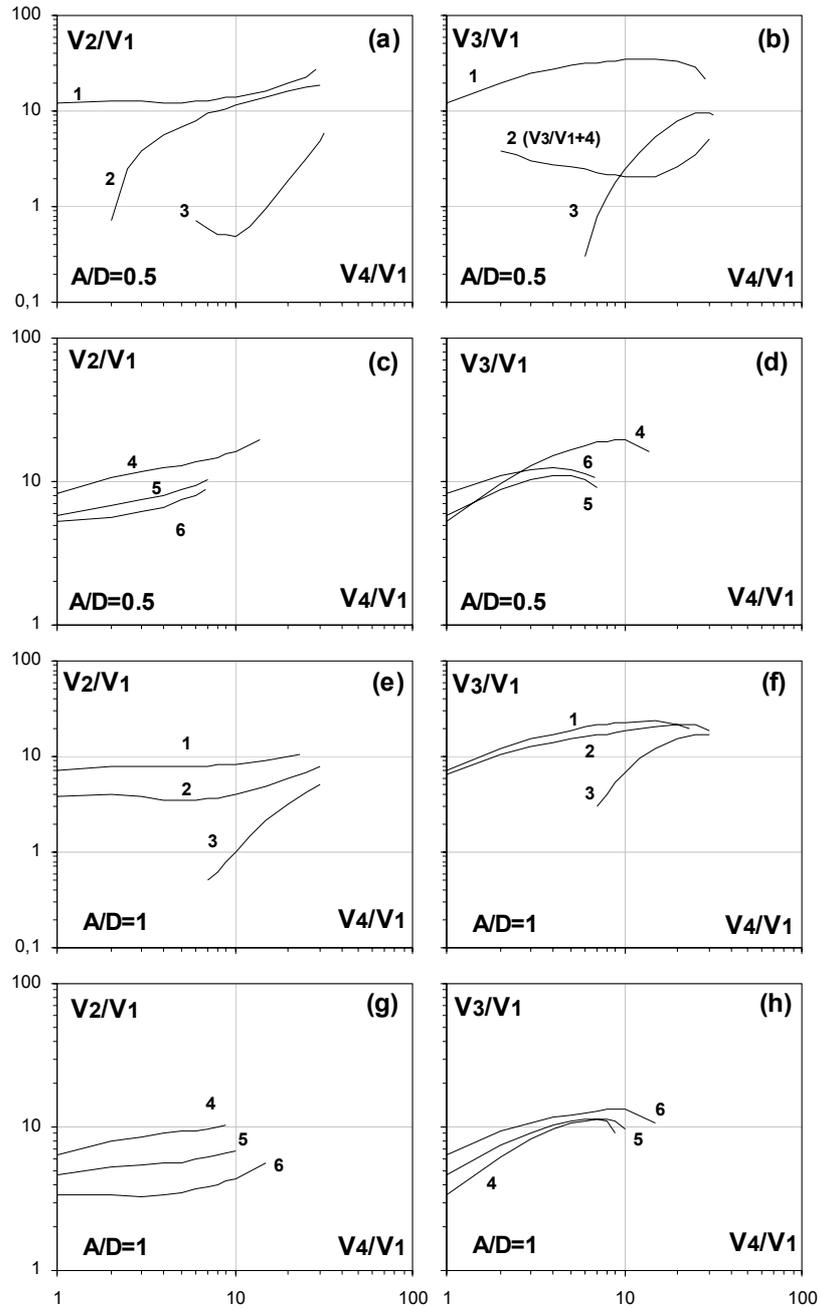

**Figure 9.** The relationship between voltage ratios $V_2/V_1$ and $V_3/V_1$ for A/D=0.5 (a)-(d) and 1 (e)-(h) as a function of $V_4/V_1$ for six combinations of P-Q-M specified in table 1.

As can be concluded from figure 9, the range of ratio $V_4/V_1$ is limited for a given P, Q and M values. Beyond this limit of $V_4/V_1$ a focusing condition cannot be achieved for any pair of voltages $V_2/V_1$ and $V_3/V_1$ and focus point lies outside the image plane. For P2Q3M1 and P3Q4M1 (defined number 2) the voltage ratios $V_2/V_1$ and $V_3/V_1$ change in an opposite way for two lens systems A/D=0.5 and 1. Because for A/D=0.5 the ratios of $V_2/V_1$ and $V_3/V_1$ have negative values to provide the focus condition with constant M. The range of the values of $V_4/V_1$ for A/D=1 includes $V_4/V_1$=1.0. Other combinations are very similar.

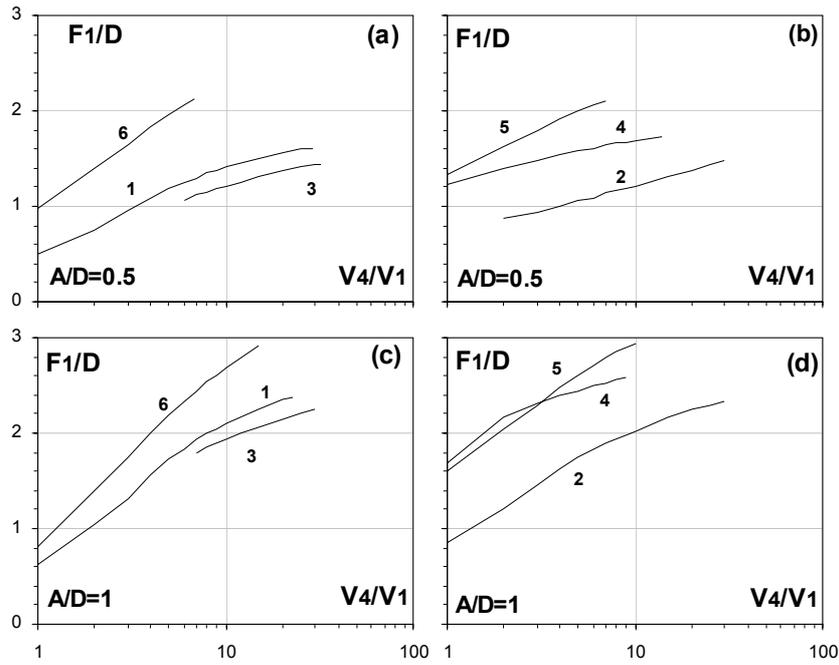

**Figure 10.** Values of the first mid-focal lengths $F_1/D$ for four-element lenses with $A/D=0.5$ (a-b) and 1 (c-d) as a function of $V_4/V_1$ for six combinations of P, Q and M.

We show the first mid-focal length $F_1/D$ as a function of $V_4/V_1$ in figure 10 for six combinations of $A/D=0.5$ and 1. It is important that the focal length of a lens does not depend on P and Q or M, but on cylinder lengths and potentials. However, the potentials of the lens are changed to provide constant values of P, Q and M. So, the focal lengths ($f_1$, $f_2$, $F_1$ and $F_2$) may vary if the potentials of the lens element are not constant. Other lens parameters ($F_2/D$, $f_1/D$ and $f_2/D$) can be obtained from equation (1) for a fixed value of P, Q and M. The calculated values of the spherical and chromatic aberration coefficients are shown in figure 11 as a function of $V_4/V_1$ for each of the six combinations. Under these conditions it is possible to use a four-element lens as the spherical and chromatic aberration coefficients are then smaller for a given focusing conditions.

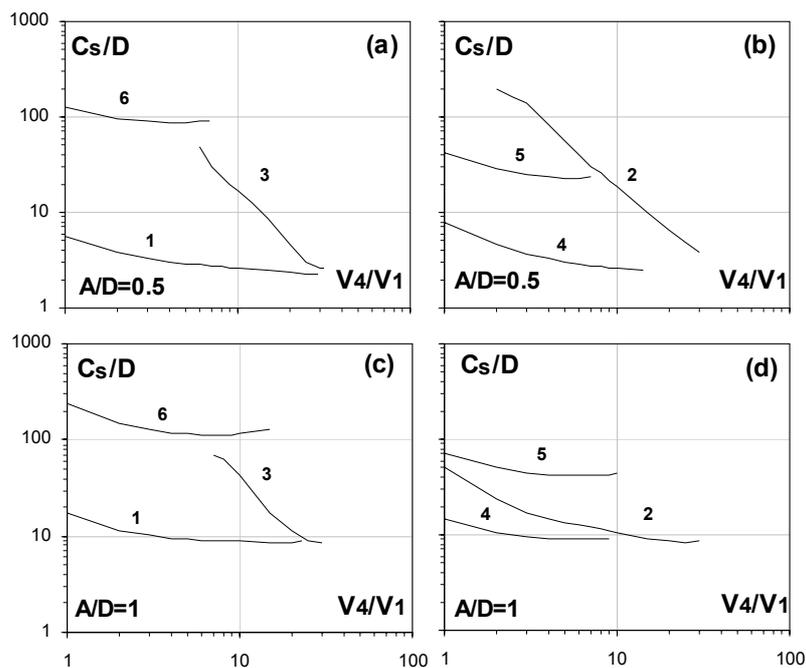

**Figure 11.** Values of the spherical aberration coefficient $C_s/D$ for four-element lenses with A/D=0.5 (a-b) and 1 (c-d) as a function of $V_4/V_1$ for six combinations of P, Q and M.

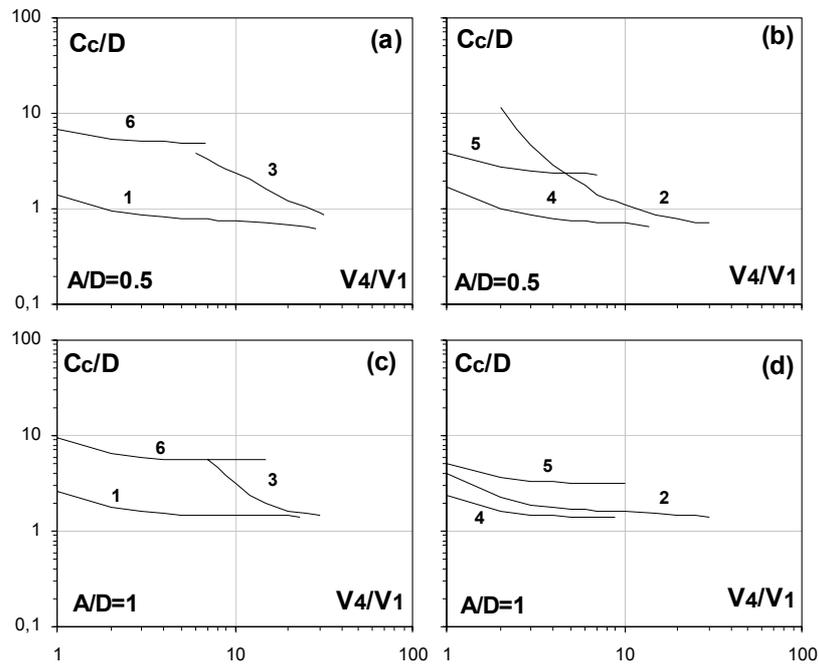

**Figure 11.** Values of the chromatic aberration coefficient $C_c/D$ for four-element lenses with A/D=0.5 (a-b) and 1 (c-d) as a function of $V_4/V_1$ for six combinations of P, Q and M.

4.3. Five-element lenses

This part is a simple analysis using LENSYS and SIMION of the standard five-element lens of total length 6 and 8 D. A schematic view of a five element lens system with the lens parameters is shown in figure 12. The unit of distance is the cylinder diameter D and the gap between two cylinders is G=0.1D. The lengths of the lens elements discussed here are given in table 2. To prevent confusion of terms we have referred to these lens systems as A/D=0.5 and 1. The dependences of voltage ratios on magnification and aberration coefficients are presented graphically for two modes of five-element lens systems, true and afocal zoom lens modes.

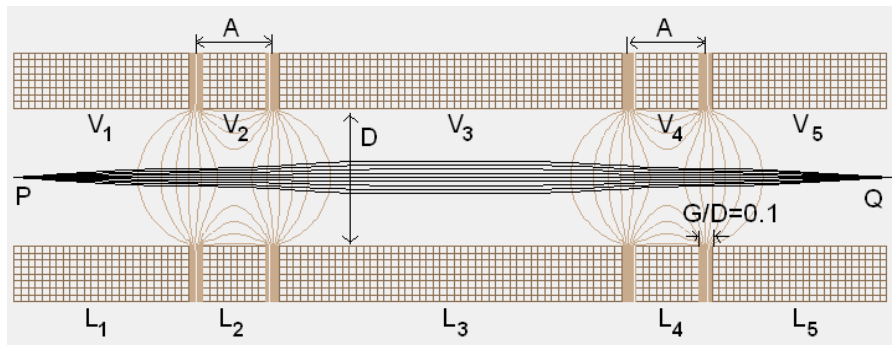

**Figure 12.** Schematic diagram of a five-element electrostatic lens system.

| Lens type | $L_1$ | $L_2$ | $L_3$ | $L_4$ | $L_5$ |
|---|---|---|---|---|---|
| A/D=0.5[*] | 1.25 | **0.5** | 2.5 | **0.5** | 1.25 |
| A/D=1[*] | 1.5 | **1** | 3 | **1** | 1.5 |
| A/D=1.5[†] | 1.5 | **1.5** | 3 | **1.5** | 1.5 |

**Table 2.** The lengths of the lens elements.

[*] These dimensions are given in LENSYS program as Varimag 6 and 8.
[†] Heddle and Papadovassilakis [47] and Trager-Cowan et al [48]

For the true zoom lenses, to form a focal point of electrons with the same starting angle, an optimized relationship between the voltages $V_2/V_1$ and $V_4/V_1$ has to be calculated for a given voltage ratio $V_5/V_1$ of final to initial electron energy. The zoom lens condition for five-element lenses having the potentials of these middle electrodes may be achieved with each of the potentials at a high and low value. It is practical importance that the cases with $V_2/V_1>1$ and $V_4/V_1>1$ have smaller aberration coefficients. Then other cases have higher spherical and chromatic aberrations [47, 48]. Some example of calculated values of $V_2/V_1$ and $V_4/V_1$ are shown with the spherical and chromatic aberration coefficients, $C_s/D$ and $C_c/D$, in figure 13 for $V_5/V_1=0.1$, 1 and 10 with $V_3/V_1= (V_5/V_1)^{1/2}$, for two types of lens system A/D=0.5 and 1. Similar curves can apply for other values of $V_5/V_1$.

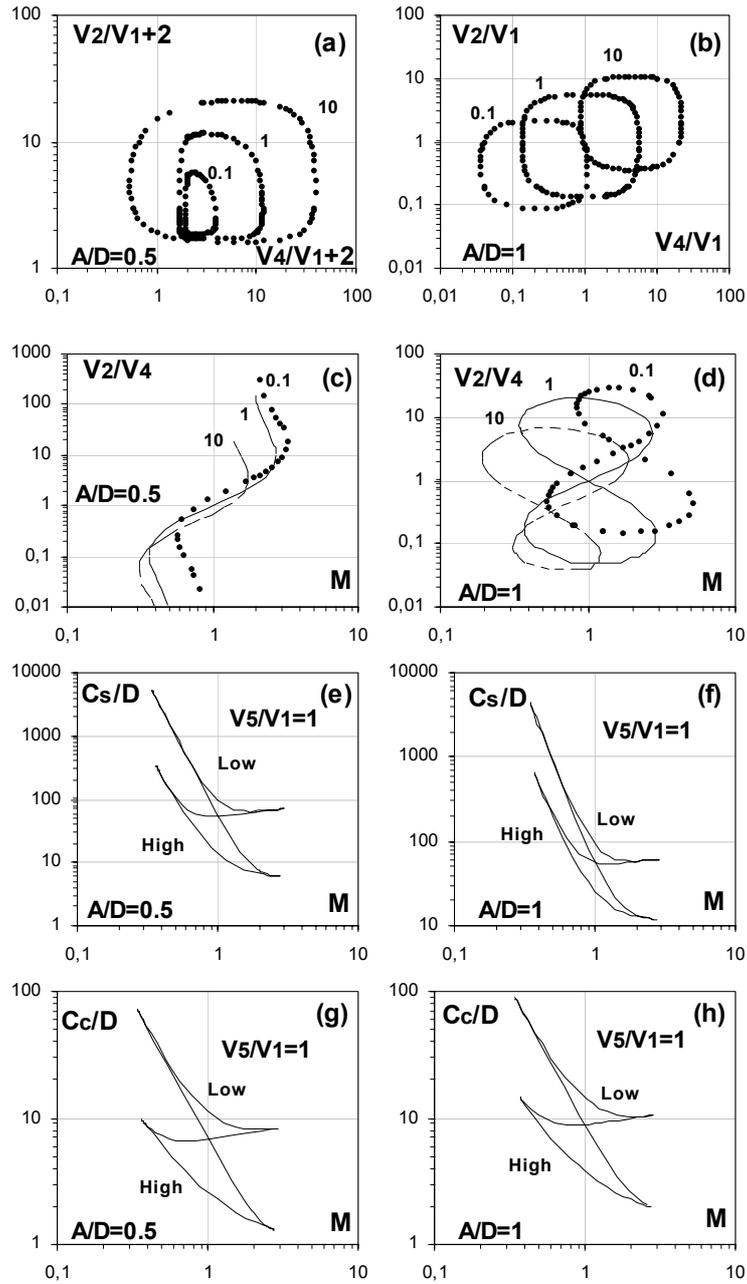

**Figure 13.** Values of $V_2/V_1$ and $V_4/V_1$, and the variation of $V_2/V_4$ which contains two voltage ratios of middle electrodes as a function of the magnification for the five element lens systems having A/D=1 and 0.5 with the spherical and chromatic aberration coefficients, $C_s/D$ and $C_c/D$, for three different values of $V_5/V_1$. Note that the scale of $V_2/V_1$ and $V_4/V_1$ is shifted two units for A/D=0.5.

The dependences of the ratio $V_2/V_4$ on the magnification for two type of lens system are shown in figure 13. It can be seen that for three value of $V_2/V_4$, there are only one point corresponding to the magnification and for other values two points corresponding to M. For a given value of $V_5/V_1$, one of these three points is the crossing point on the curve, and other points are upper and lower bound of the zoom lens. It is important that the crossing point on figure 13(d) shows the required voltage ratios for afocal lenses. In the afocal zoom lens mode to produce a final beam angle of zero at infinity, the lenses are arranged so that the second focal point of the first three-element lens ($F_2$) coincides with the first focal point of the second lens ($F_1$). Therefore the magnification is independent of the object and image positions and is determined by the overall voltage ratios. The conditions of "afocal zoom" lens are given in the following equations with $M = (V_5/V_1)^{-1/4}$.

$$\frac{V_5}{V_3} = \frac{V_3}{V_1} \qquad \frac{V_4}{V_3} = \frac{V_2}{V_1} \qquad (2)$$

We calculated the voltage ratios and magnification properties of two types of lens system for $V_5/V_1=0.1$, 1 and 10. For other values of $V_5/V_1$ we can determine the necessary values of $V_2/V_1$ and $V_4/V_1$ with the constant magnification. Therefore five-element lens system can be operated to give constant magnification overall voltage ratio. By a simple calculation of the various voltage ratios and magnification, a diagram showing that the lines of the constant parameters can be generated, where we used the notation of Ref. [47], and additionally with aberration effect. In this way figure 15 illustrate the calculated values of voltage ratios and spherical aberration coefficient in terms of M and $V_5/V_1$, for two types of lens systems. The range of the magnification is from 0.4 to 3, and for $V_5/V_1$ it is from 0.04 to 70. These conditions are especially convenient for low and high energy electron spectroscopy.

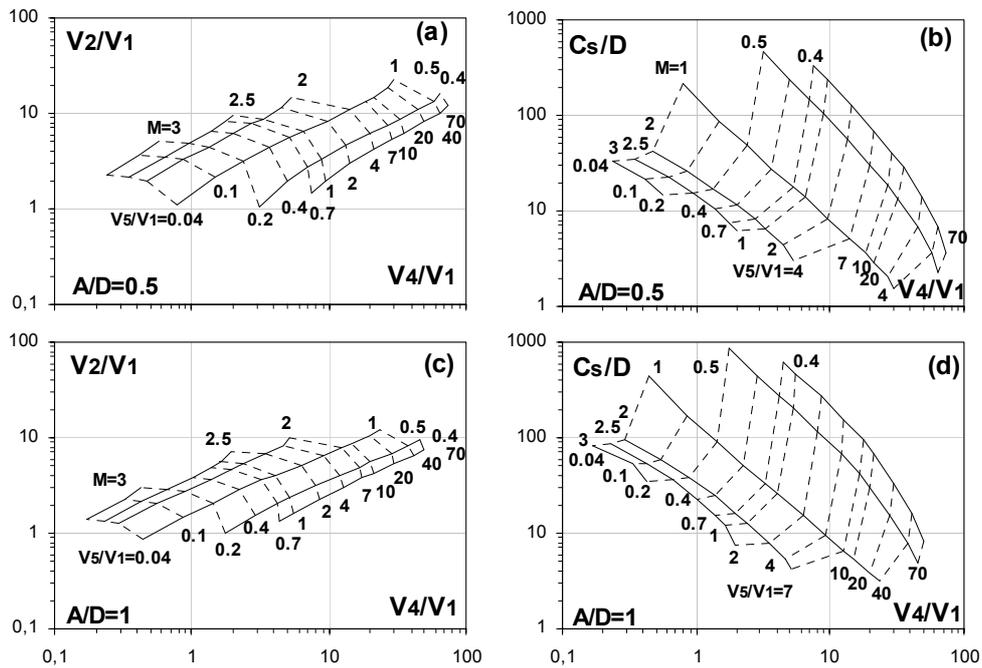

**Figure 15.** Lines of constant magnification and overall voltage ratio for two different type of five element lens. Values of $V_2/V_1$, $V_4/V_1$ and $C_s/D$ can be read off for any chosen combination of these parameters.

## 5. Conclusions

In this paper we reviewed the properties of multi-element cylinder lenses. We started with a discussion of the criteria for choosing the most appropriate dimensions of the lens elements for the lens systems, and then presented results of three-element lenses with the zoom-lens and acceleration/deceleration-lens mode. The required dependencies of the voltage ratios and aberration coefficients of resulting zoom lens are presented. Therefore the characteristics of the three-element lens have been determined.

Secondly, two types of four-element lens systems have been investigated in great detail. Focal parameters, zoom lens properties and aberration coefficients have been calculated for six different configurations of these lens systems. It is shown that a higher value of $V_2/V_1$ and $V_3/V_1$ has the lowest spherical and chromatic aberration coefficients and other cases have higher spherical and chromatic aberration coefficients, and additionally the greater the object distance P the wider the aberration coefficients for a given ratio $V_4/V_1$ allowing a focal point at the image plane. A few of the obtained results are compared to the previous studies [36, 37] and a very good agreement with the previous data was achieved.

Finally, the five-element lens systems have been investigated with true and afocal zoom lens modes, and the required voltage ratios and magnification of lens systems have been performed over a wide energy range. It is shown that the range of magnification of any given four-element lens is quite limited when it is compared with the five-element afocal case. An important conclusion from the calculations the computer based techniques can be used for accurate modelling of multi-element lens systems.

**Acknowledgement**

This work is supported by Government Planning Organisation, through grant 2002K120110 and Afyon Kocatepe University, Scientific Research Project Commission, through grant 031-FENED-07.